\documentclass[aps,pra,reprint,amsmath,amssymb,superscriptaddress]{revtex4-1}

\pdfoutput=1
\usepackage[T1]{fontenc}
\usepackage[latin9]{inputenc}
\usepackage[english]{babel} 
\usepackage{amsmath}
\usepackage{color}
\usepackage{physics}
\usepackage{graphicx}
\usepackage{dsfont}

\makeatletter
\@ifundefined{textcolor}{}
{
 \definecolor{BLACK}{gray}{0}
 \definecolor{WHITE}{gray}{1}
 \definecolor{RED}{rgb}{1,0,0}
 \definecolor{GREEN}{rgb}{0,1,0}
 \definecolor{BLUE}{rgb}{0,0,1}
 \definecolor{CYAN}{cmyk}{1,0,0,0}
 \definecolor{MAGENTA}{cmyk}{0,1,0,0}
 \definecolor{YELLOW}{cmyk}{0,0,1,0}
}

\newtheorem{theorem}{Theorem}

\newtheorem{definition}{Definition}

\makeatother

\usepackage{babel}
\begin{document}

\title{Gauge-invariant quantum thermodynamics: consequences for the first law}

\author{Lucas C. C\'{e}leri\href{https://orcid.org/0000-0001-5120-8176}{\includegraphics[scale=0.05]{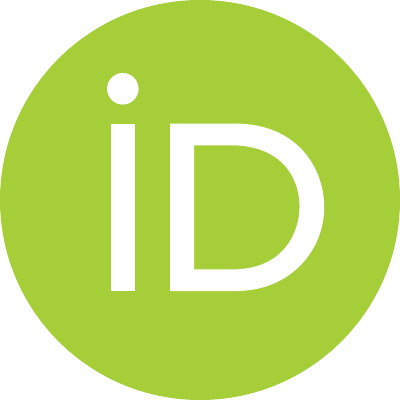}}}
\email{lucas@qpequi.com}
\affiliation{QPequi Group, Institute of Physics, Federal University of Goi\'{a}s, 74.690-900, Goi\^{a}nia, Brazil}

\author{\L ukasz Rudnicki}
\email{lukasz.rudnicki@ug.edu.pl}
\affiliation{International Centre for Theory of Quantum Technologies (ICTQT),
University of Gda\'{n}sk, 80-308 Gda\'{n}sk, Poland}

\begin{abstract}
Universality of classical thermodynamics rests on the central limit theorem, due to which, measurements of thermal fluctuations are unable to reveal detailed information regarding the microscopic structure of a macroscopic body. When small systems are considered and fluctuations become important, thermodynamic quantities can be understood in the context of classical stochastic mechanics. A fundamental assumption behind thermodynamics is therefore that of coarse-graning, which stems from a substantial lack of control over all degrees of freedom. However, when quantum systems are concerned, one claims a high level of control. As a consequence, information theory plays a major role in the identification of thermodynamic functions. Here, drawing from the concept of gauge symmetry, essential in all modern physical theories, we put forward a new possible, intermediate route. Working within the realm of quantum thermodynamics we explicitly construct physically motivated gauge transformations which encode a gentle variant of coarse-graining behind thermodynamics. As a consequence, we reinterpret quantum work and heat, as well as the role of quantum coherence.
\end{abstract}

\maketitle

%%%%%%%%%%%%%%%%%%%%%%%%%%%%%%%%%%%%%%%%%%
\section{Introduction} 

Classical thermodynamics and quantum thermodynamics rely on very different paradigms. Classical thermodynamics is based on a fundamental fact that we do not have access to microscopic degrees of freedom of a macroscopic system. Measurements reveal an average over space and time since our clocks and rules are not able to follow the underlying complex dynamics of the system. This is referred to here as low ability of controlling the system at hand, also known as coarse graining, and~it constitutes the well-known thermodynamic limit from which the thermodynamic variables emerge~\cite{Callen1985}. Leaving such limit and diving towards smaller and smaller scales where fluctuations become important, we still expect classical thermodynamics to hold on average, and~statistical mechanics can be employed in order to properly describe the system~\cite{Sekimoto2010,Seifert2012,Jarzynski1997,Crooks1999}.

Diving even deeper, we enter the realm of quantum mechanics, where thermodynamics acquires a different flavor. Here, the~focus is often on systems with a few degrees of freedom, subject to a high level of control. By~this we mean that in order to compute or measure thermodynamic quantities in this regime we do need to know the state of the system and, most of the times, also the state of the environment. In~contrast to classical thermodynamics, we can perform sharp measurements in the quantum realm. It is under this context that information theory starts playing a crucial role~\cite{Horodecki2003,Skrzypczyk2014,Bera2017,Goold2016} in the sense that thermodynamic quantities are operationally defined in terms of informational ones~\cite{Goold2016}.

\begin{figure}[h]
    \includegraphics[width=0.5\textwidth]{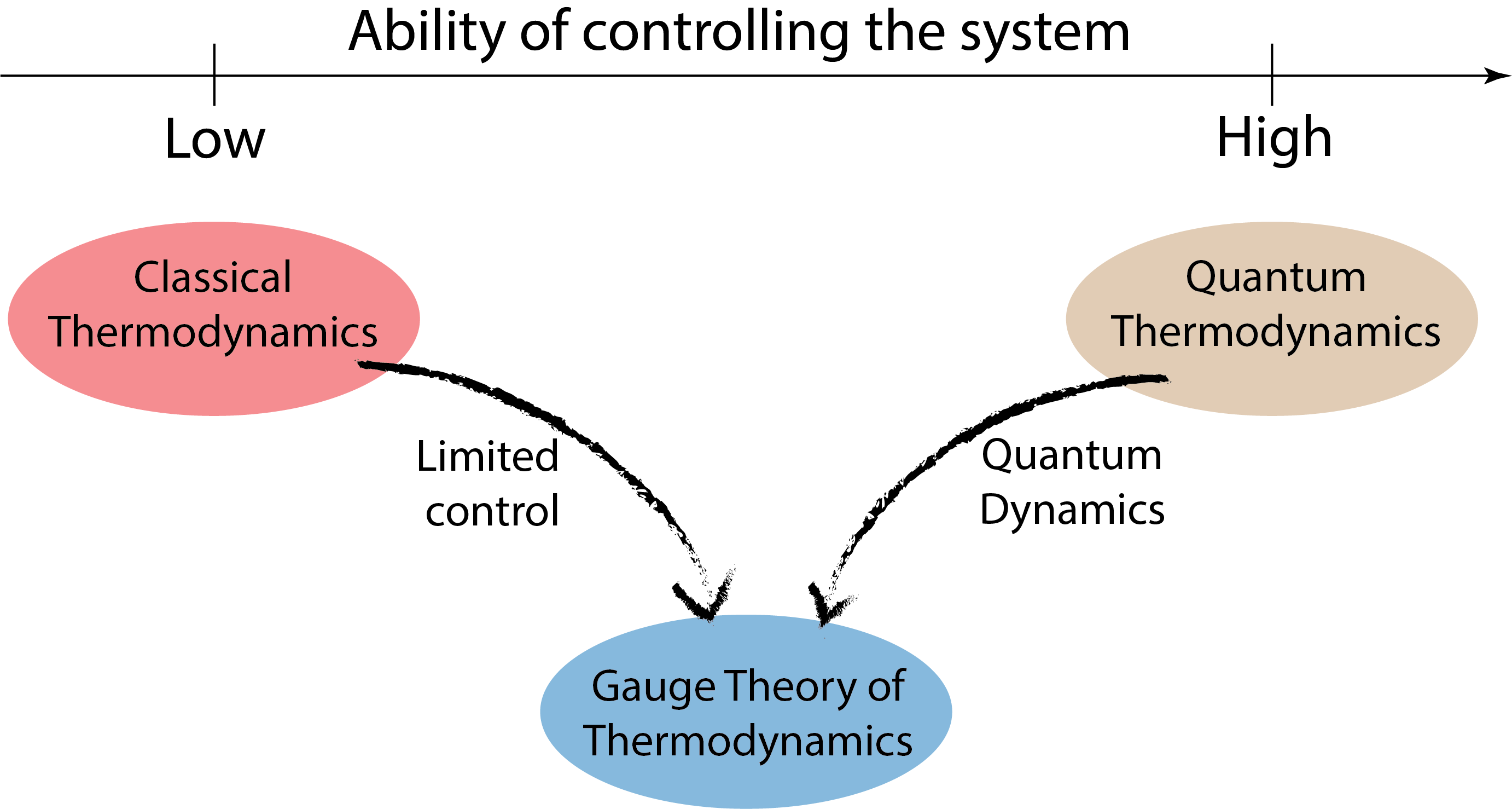}
    \caption{Emergent gauge theory of quantum thermodynamics. Quantum description of systems and their dynamics, taken together with the limited access to the details, characteristic to thermodynamics, do form an intermediate theory: quantum thermodynamics subject to an emergent gauge symmetry. Note that full quantum description is richer than full classical description. This means that while for the classical description it seems enough to employ averaging, for~the quantum description it is not sufficient. This is because in quantum physics we also carry information about bases of the Hilbert space, in~particular, the~eigenbasis of the Hamiltonian. That is the reason we do need something more than just an average process to wash away all the redundancy of the quantum description. Importantly, this redundancy is truly redundant only while discussed from a more classical point of view on thermodynamics. From~the information theory point of view (quantum dynamics, e.g., Schr\"odinger equation), this is not redundant at all.}
    \label{fig:control}
\end{figure}

Here, we propose a new route towards quantum thermodynamics, which at the same time aims to keep the spirit of classical thermodynamics. From~the point of view of thermodynamics, the~quantum state caries too much information, introducing a redundancy that should be eliminated, in~the same spirit that in classical thermodynamics, the~complete knowledge of the positions and velocities of all particles constituting the system is redundant and washed away by the average process discussed above (see Figure~\ref{fig:control} for an illustration). There is a fundamental principle in physics, named gauge invariance, that does precisely this~\cite{Cabibbo2017}. The~basic idea behind this principle is that physical quantities are invariant under certain symmetry transformations of the~system. 
%%%%%%%%%%%%%%%%%%%%%%%%%%%%%%%%%%%%%%%%%%%%%%%%%%%%%%%
%%%%%%%%%%%%%%%%%%%%%%%%%%%%%%%%%%%%%%%%%%%%%%%%%%%%%%%

The main goal of the present work is to build a gauge group for quantum thermodynamics that is able to remove the aforementioned redundant information stored in a quantum system and then define the relevant physical quantities, like work and heat, based on the gauge invariance principle. In~other words, we propose a new framework, based on the gauge principle, from~which quantum thermodynamic quantities~emerge.

\begin{figure}[h]
    \includegraphics[width=0.5\textwidth]{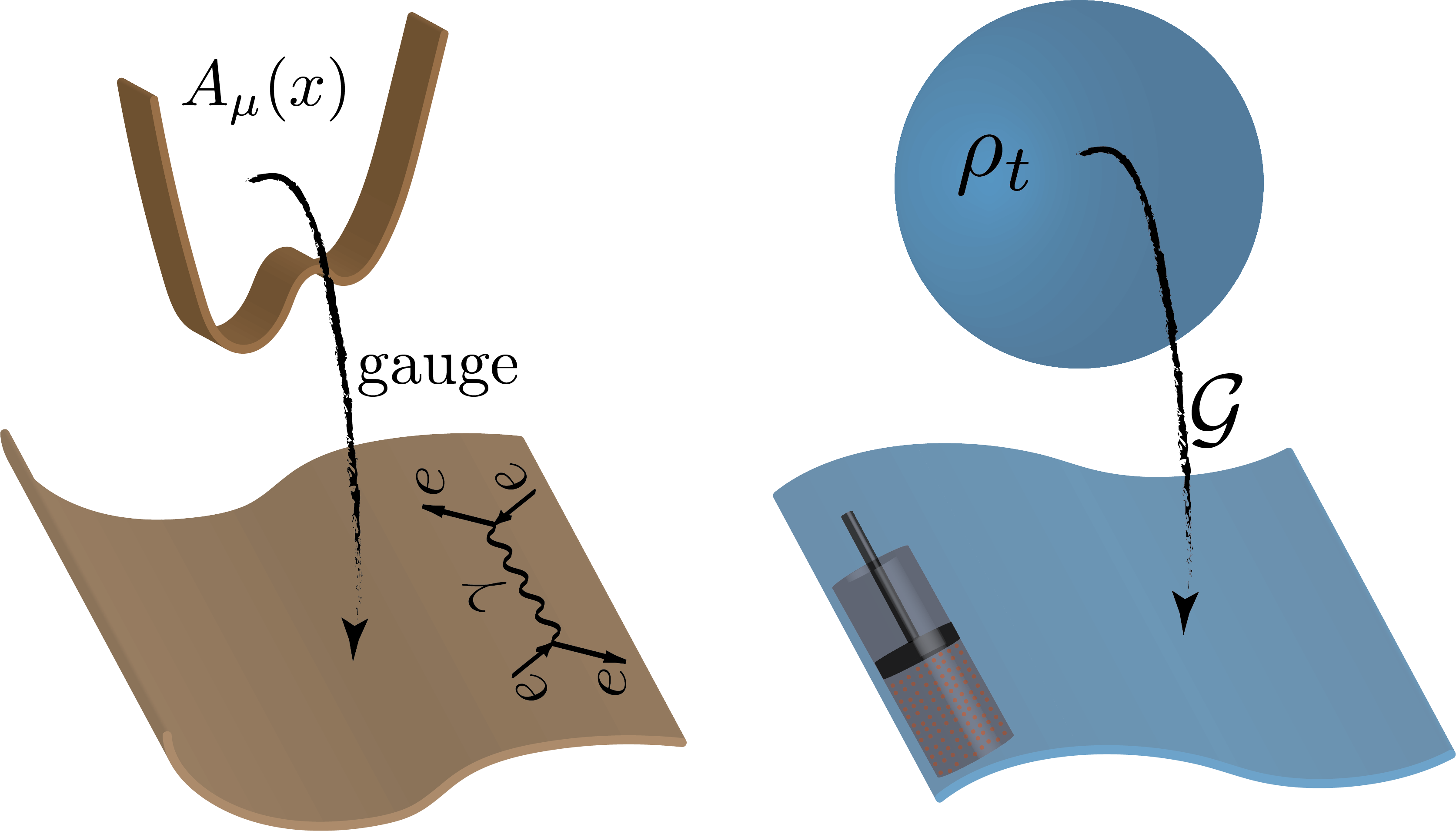}
    \caption{Emergent thermodynamic gauge. The~panel on the left illustrates the fundamental gauge behind field theories, both classical and quantum. The~set of potentials $A_{\mu}(x)$ does not have any physical meaning and the gauge transformations take us from this set to the scattering amplitudes that can be experimentally investigated. The~right panel represents the emergent thermodynamic gauge $\mathcal{G}$ proposed in this work. Such transformations take us from the set of time-dependent information carriers $\rho_{t}$, that convey no thermodynamic meaning, to~the set of thermodynamically meaningful quantities. The~gauge emerges from the coarse-graining nature of thermodynamics, being, in~this sense, not fundamental as the one behind field theories. In~other words, the~information here is only redundant from the thermodynamic point of view, in~the same sense that the positions and velocities of individual molecules of a gas are redundant in the context of classical~thermodynamics.}
    \label{fig:gauge}
\end{figure}

It is important to observe that there is a crucial difference between the gauge symmetry of classical and quantum field theories, and~the one we identify behind thermodynamics. In~the former case, the {\textit{fundamental}} gauge transformations take us from potentials (e.g., vector potential in electrodynamics) to physically meaningful quantities such as scattering amplitudes~\cite{Cabibbo2017}. As~already mentioned, in~our theory, the~gauge takes us from information theory to the thermodynamic quantities. In~other words, the~principle of gauge invariance is employed here exactly to wash away information which is ``too detailed''. However, this redundancy is not fundamental, as~in the case of classical and quantum field theories, but~is redundant only in the context of thermodynamics and not, for~instance, in~the context of information theory. That is why we speak about an \textit{{emergent}} gauge theory. Figure~\ref{fig:gauge} illustrates this~idea.

%%%%%%%%%%%%%%%%%%%%%%%%%%%%%%%%%%%%%%%%%%%%%%%%%%%%%%%
%%%%%%%%%%%%%%%%%%%%%%%%%%%%%%%%%%%%%%%%%%%%%%%%%%%%%%%

Although our theory is built with reference to mean energy, which is always unambiguously defined, it provides a general framework to define every thermodynamic quantity. As~an application, we infer conclusions relevant for the first law of quantum thermodynamics. In~particular, we construct notions of invariant work and heat. Moreover, our framework allows us to clarify the role of quantum coherences, because~the resulting gauge-invariant notion of heat covers delocalized energy, thus being present even in closed quantum systems. Note that in~classical thermodynamics, when a system is taken out of equilibrium, energy fluxes (in~the form of heat) can be generated inside the system. Thus, the~concept of heat in closed classical systems is well-known~\cite{Callen1985}, even in relativistic systems~\cite{Einstein}. In~the context of quantum thermodynamics, the~notion of heat in closed systems contains some subtleties ---as well as the notion of work--- but it is not a new issue and has been considered in terms of the generation of quantum coherences in the energy eigenbasis (see Refs.~\cite{Bunin2011,Polkovnikov2008,Lostaglio2015,Kammerlander2016}, just to mention a few). However, it is important to observe here that our theory goes far beyond such results. It is just a coincidence that these earlier definitions agree with our theory. Actually, as~such notions naturally emerge from the gauge group proposed here, our results put these earlier definitions under the umbrella of gauge symmetry. Note that the more frequently adopted approach, in~which heat does not exist in closed quantum systems, does not agree with our theory in general, as~explained in detail in the~text. 

The paper is organized as follows. The~next section is devoted to the definition of the emergent thermodynamic gauge, where it is explicitly built and motivated. Based on the gauge invariance principle, the~new definitions of average work and heat are presented in Sections~\ref{sec:work} and~\ref{sec:heat}, respectively. After~presenting illustrative examples in Section~\ref{sec:examples}, we finish with a discussion in Section~\ref{sec:discussions}, where we also outline some consequences of our approach for the statistical definitions of work and~heat.

%%%%%%%%%%%%%%%%%%%%%%%%%%%%%%%%%%%%%%%%%%%%%%%%%%%%%%
%%%%%%%%%%%%%%%%%%%%%%%%%%%%%%%%%%%%%%%%%%%%%%%%%%%%%%
%%%%%%%%%%%%%%%%%%%%%%%%%%%%%%%%%%%%%%%%%%%%%%%%%%%%%%
\section{Emergent Thermodynamic~Gauge} 
\label{sec:gauge}

In field theories, the~potentials do not have physical meaning in the sense that they cannot be measured. The~gauge transformations enter here as fundamental symmetries of the system, coming from the redundancy in the description of the relevant degrees of freedom. Considering thermodynamics, carriers of quantum information---the physical states of the system---play the same role as the potentials do, while thermodynamic quantities are considered as the physically relevant ones. The~gauge, in~this sense, is not fundamental since all thermodynamic quantities emerge from a microscopic description by a process of coarse graining. The~gauge emerges from this process. At~the same time, the~microscopic description is physically meaningful, and~experimentally accessible. Though, not from the perspective of standard~thermodynamics. 

The main goal of this section is to propose a gauge formalism that is able to extract relevant information from a quantum state of the system. Relevant only in the context of thermodynamics. Therefore, the~primary connection between the gauge theory proposed here and the one in field theories is that in both cases, the~gauge invariance principle is employed to eliminate redundant~information.

To be precise, let $H_t$ be a (possibly time-dependent) Hamiltonian of our system, and~let $\rho_t$ be a density matrix describing it, which, considered in its entity, is here ascribed to the domain of information theory. In~the present analysis, we restrict the attention to these two fundamental objects, however, the~developed formalism can easily be extended to include more observables, such as the total number of~particles.

In a thermodynamic, coarse-grained description, we are only interested in a particular set of quantities, such as the mean energy which is essential for the first law of thermodynamics
\begin{equation}\label{eq8}
    U\left[\rho_t\right]=\textrm{Tr}\left(\rho_t H_t\right).
\end{equation}
In our notation, the~mean energy is a functional depending on the density matrix, with~the latter being a function of~time.

The next crucial quantity is work associated with a given process that takes place during the time interval $t\in[0,\tau]$, defined as~\cite{Alicki1979}
\begin{equation}
W_{u}\left[\rho_t\right]=\int_{0}^{\tau}dt\textrm{Tr}\left(\rho_{t}\dv{H_{t}}{t}\right).
\label{Work not GI}
\end{equation}
We add a subscript ``u'', derived from a word ``usual'', to~emphasize the fact that we start the discussion with standard notions met in quantum thermodynamics. Later on, we construct gauge transformations which lead us to the corresponding invariant quantities, later being denoted with the subscript~``inv''.

From now one we shall use the dot to denote time derivatives, so that the integrand in (\ref{Work not GI}) becomes $\textrm{Tr}\left(\rho_{t}\dot H_{t}\right)$. The~functional $W_u$, even though we do not write that explicitly, depends on the time derivative of the Hamiltonian. We omit this dependence because the Hamiltonian does not belong to the set of dependent variables of the~theory.

Furthermore, we consider an associated functional
\begin{equation}\label{heatusual}
Q_u\left[\rho_t\right]=\int_{0}^{\tau}dt\textrm{Tr}\left(\dot\rho_{t} H_{t}\right).
\end{equation}
which is the usual notion of heat~\cite{Alicki1979,Kieu2004}. This functional apparently depends on the time derivative of the density matrix. However, due to energy conservation $Q_u=U-W_u$, the~heat defined in (\ref{heatusual}) can be expressed solely by $\rho_t$.

In fundamental physics, gauge is realized as an action of a gauge group. Elements of this group act on variables describing a given theory (potentials). In~our parallel, the~role of the potentials is played by the carriers of quantum information, namely, the~density matrix $\rho_t$. Therefore, emergent gauge transformations need to be represented as (some) unitary matrices $V_t$ acting as follows
\begin{equation}\label{transfor}
    \rho_t \mapsto V_t \rho_t V_t^\dagger.
\end{equation} 

The emergent gauge is allowed to be time-dependent, the~same way as the standard one in field theories depends on space and time. As~a consequence, the~time derivative of the density matrix transforms as
\begin{equation}
\dot\rho_t \mapsto V_t \dot\rho_t V_t^\dagger+\dot V_t \rho_t V_t^\dagger+V_t \rho_t \dot V_t^\dagger,
\end{equation}
again resembling familiar transformation~patterns.

From a more operational perspective we can think about the transformation (\ref{transfor}) as about a choice of a (time-dependent) basis in which we wish to describe the state of the system. On~the other hand, the~eigenbasis of the Hamiltonian plays the role of a fixed reference frame. Therefore, a~more general transformation rule in which together with (\ref{transfor}) the Hamiltonian is being ``covariantly'' transformed as $H_t \mapsto V_t H_t V_t^\dagger$, is not the route followed here. While such a transformation not only preserves the mean energy but, more importantly, resembles Lorentz transformations in Special and diffeomorphisms in General Relativity, the~resulting theory assuming invariance with respect to it is void. This happens because the associated gauge group contains all unitary transformations, simply being too large. As~we will convince ourselves soon, the~gauge group stemming from (\ref{transfor}) is more~gentle.

We now observe that, in~field theories, only the quantities which are \textit{{gauge-invariant}} are considered as being measurable and physically meaningful. Within~ our framework of the emergent gauge in thermodynamics, we adopt the same philosophy, namely, we require thermodynamically meaningful quantities to be invariant with respect to the proposed emergent gauge defined in Equation~(\ref{transfor}). Clearly, not all unitary transformations can be allowed to render the emergent gauge transformations, as~otherwise the theory would be trivial. To~select a desired set of transformations we will build on the primary notion of our framework, namely, the~mean energy which has to be unambiguously gauge-invariant. We therefore define:

\begin{definition}[Emergent thermodynamic gauge]
Unitary transformations are admissible gauge transformations if they preserve the mean energy, independently of a particular state of the system, i.e.,
\begin{equation}
    U[V_{t}\rho_{t}V^{\dagger}_{t}] = U[\rho_{t}]
\end{equation}
for all $\rho_{t}$.
\end{definition}

Such a definition splits the state space into equivalence classes, under~the mean energy. Inside each element of this set, all the density operators lead to the same mean energy, although~they can hold very different informational meanings. That is precisely the sense in which we say that the gauge is emergent and not a fundamental one. It is based on our lack of control over the system and not on a fundamental redundancy in the description of the system. This is very natural when we talk about quantum systems since, in~general, we are only able to perform a limited set of measurements and the complete reconstruction of the density matrix is not practical. For~instance, in~the experiment described in Ref.~\cite{Ribeiro2020}, although~dealing with a closed quantum system, quantum state tomography is prohibitive due to the dimension of the Hilbert space, and~only energy measurements are allowed. If~we consider continuous quantum systems, not even sharp energy measurements are allowed and energy measurements encompass a finite size energy~window.

By virtue of the above definition we are in position to explicitly construct the gauge. The~invariance of the mean energy of any state requires that $\textrm{Tr}\left[\rho_t \left( V_t^\dagger H_t V_t-H_t\right)\right]=0$ independently of $\rho_t$, which simply means that the unitary operations need to commute with the Hamiltonian, i.e.,~$\left[V_t,H_t\right]=0=\left[V_t^{\dagger},H_t\right]$. Therefore, thermodynamic gauge transformations are simply unitary channels which commute with the Hamiltonian. It is important to observe that our emergent gauge is not equivalent to the well-known thermal operations employed in the resource theory of thermodynamics as free transformations~\mbox{\cite{Brandao2013,Chitambar2019}}. {The gauge group we built involves trivial thermal operations in which there is no interaction between the system and a potentially present bath. Consequently, the~bath can be traced out, as~it plays no role in defining the gauge process.} The emergent gauge rather forms a very particular subclass of the latter operations which, importantly, are not attributed to additional physical~interpretation.

While exploring the structure of the gauge, we first assume that we work with a $d$-dimensional system. Then, $V_t\in\mathcal{U}\left(d\right)$, with~$\mathcal{U}\left(d\right)$ representing the set of $d\times d$ unitary matrices. In~fact, unitaries which commute with $H_t$, by~definition form a subgroup of $\mathcal{U}\left(d\right)$. In~this way we again obtain a very close contact with the fundamental gauge in field theories, as~this subgroup can be called the \emph{{emergent gauge group}}. 

Now, since the Hamiltonian is Hermitian, we can conveniently parametrize all the members of the emergent gauge group. To~this end we express the Hamiltonian as
\begin{equation}
H_t=u_t\left(\bigoplus_{k=1}^{p}\varepsilon_{k}(t)\mathds{1}_{n_{k}}\right)u_t^{\dagger} \equiv u_t h_t u_{t}^{\dagger},
\label{eq:repre}
\end{equation}
where $p$ is the number of its distinct eigenvalues, $n_{k}$ is the multiplicity of each $\varepsilon_{k}(t)$, such that $\sum_{k=1}^{p}n_{k}=d$, and~$u_t\in\mathcal{U}\left(d\right)$. For~the sake of clarity, we assume that the degeneracy of the Hamiltonian does not change in time, however, below~and whenever appropriate, we pinpoint the consequences of an alternative scenario. A detailed mathematical account of the latter is left for future~work.

Every complex $d\times d$ matrix $M$ commutes with~\eqref{eq:repre} if and only if it is of the form
\begin{equation}\label{eq88}
M=u_t\left(\bigoplus_{k=1}^{p}M_{k}\right)u_t^{\dagger},
\end{equation}
where each $M_{k}$ is an $n_{k}\times n_{k}$ block with arbitrary matrix elements. The~constraint $\left[V_t,H_t\right]=0$ therefore enforces
\begin{equation}\label{unitdec}
V_t=u_t\left(\bigoplus_{k=1}^{p}v_{k}\right)u_t^{\dagger},
\end{equation}
with $v_{k}\in\mathcal{U}\left(n_{k}\right)$. From~a topological point of view, the~emergent gauge group is isomorphic to
\begin{equation}
\mathcal G = \mathcal{U}\left(n_{1}\right)\times\mathcal{U}\left(n_{2}\right)\times\ldots\times\mathcal{U}\left(n_{p}\right).
\label{eq:topology}
\end{equation}
As long as the degeneracy of $H_t$ does not change in time, the~above structure remains intact. If~this is not the case, it just means that the topological structure in Equation~\eqref{eq:topology} does change as well. Moreover, we observe that even in the non-degenerate case, i.e.,~when $p=d$ and $n_1=\cdots=n_d=1$, the~group is more than just a d-fold tensor product of trivial phases just being $U(1)$ elements. For~example, all time-dependent unitary operations generated by $H_t$ belong to $\mathcal{G}$.

On the group $\mathcal G$ there is an invariant Haar measure induced from that on $\mathcal{U}\left(d\right)$. Therefore, the~group averaging allows one to assign to each relevant quantity its gauge-invariant version:
\begin{definition} [Gauge-invariant quantities]\label{def2}
Given a functional $F\left[\rho_t\right]$, its counterpart invariant with respect to the emergent gauge is
\begin{equation}
F_\mathrm{inv}\left[\rho_t\right]= \int d\mathcal G F\left[V_t\rho_t V_t^\dagger\right],
\label{eq:average}
\end{equation}
where $V_t$ is given in~\eqref{unitdec} and $d\mathcal G$ is the Haar measure on the group $\mathcal G$ defined in Equation~\eqref{eq:topology}.
\end{definition} 
It is important to remember that the Haar measure is normalized, i.e.,~$\int d\mathcal G =1$. When the degeneracy of $H_t$ is not fixed, this normalized measure just needs to be replaced by an appropriate (still normalized) functional measure $\mathcal D \mathcal G_t$ where $\mathcal G_t$ is a counterpart of~\eqref{eq:topology}, with~all $\{n_k\}$ and $p$ being~time-dependent.

Before proceeding to a particularly relevant case of work and heat, let us point out that there is a natural class of quantities which are gauge-invariant by construction. These are all quantities solely defined on the equilibrium manifold, or~even on a family of instantaneous equilibrium operators $\rho_t^{\beta}\sim e^{-\beta H_{t}}$, with~$\beta$ being the inverse temperature. Also, other states which solely depend on the Hamiltonian, e.g.,~the microcanonical state, are~preserved.

{As a final comment, it is important to make it clear that we have two transformations here. First, we have the unitary transformations that are elements of the gauge group $\mathcal{G}$. These are only symmetry transformations, not dynamical processes. Their role is to identify the set of states, at~every instant of time, that cannot be distinguished by measuring thermodynamic variables. The~other set of transformations are the physical processes, which can be any completely positive and trace preserving map, including adiabatic (reversible) transformations. These are the ones governing the time evolution of the physical system. Therefore, our theory applies to both equilibrium and non-equilibrium transformations.}

%%%%%%%%%%%%%%%%%%%%%%%%%%%%%%%%%%%%%%%%%%%%%%%%%
%%%%%%%%%%%%%%%%%%%%%%%%%%%%%%%%%%%%%%%%%%%%%%%%%
%%%%%%%%%%%%%%%%%%%%%%%%%%%%%%%%%%%%%%%%%%%%%%%%%
\section{Gauge-Invariant~Work} 
\label{sec:work}

It is an easy task to recognize that the average work defined in~\eqref{Work not GI} transforms as 
\[
W_u[\rho_t]\mapsto W_u[\rho_t]+\int_{0}^{\tau}dt\textrm{Tr}\left[H_{t},\rho_{t}\right]V_{t}^{\dagger}\dot{V}_{t},
\]
with the last term being not invariant with respect to the emergent gauge. This transformation rule has a deep physical meaning, as~the work functional defined in Equation~(\ref{Work not GI}) turns out to be gauge-invariant only when $\left[H_{t},\rho_{t}\right]=0$ for all $t$. The~latter means that the process is adiabatic in the sense that no coherences are generated during time evolution (no transitions between energy eigenstates are allowed). Such a restriction is required since we want the invariance to hold for all choices of $V_t\in\mathcal{G}$. It is interesting that such a notion of adiabatic transformation was already considered in literature~\cite{Bunin2011,Polkovnikov2008}. We will comment more on this when discussing the invariant~heat.

However, the~general formalism presented in the preceding section allows us to provide the notion of work which is invariant with respect to the emergent gauge. 
\begin{theorem}[Gauge-invariant work]\label{th1} Let $h_{t} = u_{t}^{\dagger} H_{t} u_{t}$ be the diagonal form of the Hamiltonian $H_{t}$, with~$u_t$ unitary. Then the notion of work which is invariant with respect to the emergent thermodynamic gauge is given by
\begin{equation}
W_{\mathrm{inv}}[\rho_t]=\int_{0}^{\tau}dt\mathrm{Tr}\left[\rho_{t}u_{t}\dot{h}_{t}u_{t}^{\dagger}\right].
\label{eq:inv_work}
\end{equation}
\end{theorem}
It is straightforward to recognize that the right hand side is indeed invariant. Nevertheless, we make a formal proof in the Appendix \ref{app:invariance}. Importantly, the~above result does also hold if the degeneracy of $H_t$ changes in time, as~potential contribution coming from that fact is of zero measure with respect to the time integral. It is a bit more involved to show that the invariant notion of work is the same as would be given by Equation~\eqref{eq:average}. This fact is also demonstrated in the Appendix \ref{app:invariance}, with~the help of the group averaging~techniques.

In order to make closer contact with the usual gauge theory, let us introduce a Hermitian potential $A_{t}$ defined by the relation $A_{t}=i \dot{u}_{t}u_{t}^{\dagger}$. Using this definition, we can rewrite the invariant work as
\begin{equation}
    W_{\mathrm{inv}}[\rho_t] = \int_{0}^{\tau}\dd t \mathrm{Tr}\left[\rho_{t}D_{t}H_{t}\right],
\end{equation}
with $D_{t} \cdot = \partial_{t}\cdot + i\left[A_{t},\cdot\right]$ being a covariant derivative. Note that we have the same functional form of the usual work defined in Equation~\eqref{Work not GI} but with the covariant derivative instead of the regular one. Our covariant derivative transforms as
\begin{equation}
D_{t}\cdot\mapsto D_{t}'\cdot=\partial_{t}\cdot+i\left[A_{t}',\cdot\right],
\end{equation}
where
\begin{equation}
A_{t}'=V_{t}^{\dagger}A_{t}V_{t}-iV_{t}^{\dagger}\dot{V}_{t},
\end{equation}
can be recognized to be the standard non-abelian gauge transformation
of the potential. Note that since the notion of work obtained in Theorem
1 is invariant, we obtain the relation
\begin{equation}
\int_{0}^{\tau}dt\textrm{Tr}\left[\rho_{t}D_{t}'H_{t}\right]=\int_{0}^{\tau}dt\textrm{Tr}\left[\rho_{t}D_{t}H_{t}\right].
\end{equation}

%%%%%%%%%%%%%%%%%%%%%%%%%%%%%%%%%%%%%%%%%%%%%%%%%
%%%%%%%%%%%%%%%%%%%%%%%%%%%%%%%%%%%%%%%%%%%%%%%%%
%%%%%%%%%%%%%%%%%%%%%%%%%%%%%%%%%%%%%%%%%%%%%%%%%
\section{Gauge-Invariant~Heat} 
\label{sec:heat}

Taking the invariant notion of work together with primordial invariance of the energy, the~gauge-invariant notion of heat $Q_{\mathrm{inv}}$ immediately follows from energy conservation. The~explicit form of the heat is given below, after~the role of quantum coherence is discussed in light of our~findings.

We represent the invariant heat as consisting of two independent contributions, $Q_{\mathrm{inv}}=Q_{u}+Q_{c}$, where the usual heat has already been defined in (\ref{heatusual}), while the \textit{{coherent heat}} is now defined as
\begin{equation}\label{CH}
Q_{c}=\int_{0}^{\tau}\dd t\textrm{Tr}\left[\rho_{t}\dot{u}_{t}h_{t}u_{t}^{\dagger}+\rho_{t}u_{t}h_{t}\dot{u}_{t}^{\dagger}\right]
\end{equation} 

The quantity $Q_{u}$ has very frequently appeared in the literature as a natural definition of heat~\cite{Alicki1979,Kieu2004,Binder2018,Landi2020}. It is important to note that this term is zero for any closed evolution since $\textrm{Tr}\left[\dot{\rho}_{t}H_{t}\right] = 0$ via the von Neumann equation. Therefore, such a definition of heat is associated only with the energy exchanged between the system under consideration and some other system, like and environment for instance. However, as~mentioned earlier, thermodynamics is a theory about closed systems and well-defined notions of heat in closed systems are established both for classical and quantum systems. While in classical systems heat is associated with the energy transferred to the fast oscillating modes of the system (in a normal mode description)~\cite{Callen1985}, in~quantum systems, it is linked with the uncertainty in the energy measurement, which is caused by the generations of quantum coherences in the energy eigenbasis~\cite{Bunin2011,Polkovnikov2008}. Therefore, adiabatic transformations are those that do not allow for the generations of coherences, implying $Q_{\mathrm{inv}}=0$ and all the energy transferred to the system being equal to the invariant notion of work. In~this case, the~usual notion of work clearly coincides with the invariant one proposed~here. 

Indeed, the~additional term $Q_{c}$, the~presence of which distinguishes the invariant heat with respect to the usual one, contains only contributions from the coherences. This fact can be seen after writing it in the energy~eigenbasis. 

Let $\left\lbrace\left|a_{k}\left(t\right)\right\rangle\right\rbrace$ and $\left\lbrace\varepsilon_{k}\right\rbrace$ be the set of time-dependent energy eigenstates and eigenvalues, respectively. For~simplicity, we for a moment assume that the Hamiltonian is non-degenerate, i.e.,~all $n_k=1$. Qualitatively, however, the same results can be obtained without this assumption. Since we are talking about work, heat and thermodynamics in general, the~energy eigenbasis seems to be the most appropriate one. At~any instant of time, the~density matrix of the system can be represented in this basis as $\rho_{t}=\sum_{jl}c_{jl}\left(t\right)\left|a_{j}\left(t\right)\right\rangle \left\langle a_{l}\left(t\right)\right|$.

In Appendix \ref{app:invariance} we show that $Q_{c}$ depends only on the differences $\varepsilon_{j}\left(t\right) - \varepsilon_{k}\left(t\right)$ and, consequently, all the terms containing the diagonal coefficients of the density matrix, $c_{ii}(t)$, do vanish. This is a universal result stating that the development of internal coherences are part of the heat. This result provides a gauge flavor to earlier definitions of heat in closed quantum systems~\cite{Bunin2011,Polkovnikov2008,Lostaglio2015,Kammerlander2016}. It is important to mention here that, although~these studies consider heat and work in the context of quantum coherences, their definitions are not mathematically or physically equivalent to ours. All of these references considered distinct set of transformations or distinct physical motivations in order to define their notions of thermodynamic quantities. Our work builds a symmetry group that is able to implement a sort of (information) coarse-graining procedure from which the thermodynamic quantities uniquely~emerge.

In terms of the covariant derivative introduced earlier, we can rewrite the heat simply as
\begin{equation}
Q_{\mathrm{inv}} = \int_{0}^{\tau}\dd t\textrm{Tr}\left[H_{t}D_{t}\rho_{t}\right],  
\end{equation}
which, again, takes the same form as the usual definition of heat, but~with the covariant derivative replacing the standard~one.

It is interesting to observe that work and (total) heat are independent of the coherences in the energy eigenbasis, contrary to the standard quantum definitions. The~average total energy can be written as $U\left[\rho_{t}\right]=\sum_{k}c_{kk}\left(t\right)\varepsilon_{k}\left(t\right)$, while the invariant work takes the form $W_{\mathrm{inv}} =\int_{0}^{\tau}\dd t\sum_{k}c_{kk}\left(t\right)\dot{\varepsilon}_{k}\left(t\right)$ and the heat is written as $Q_{\mathrm{inv}}=\sum_{k}\dot{c}_{kk}\left(t\right)\varepsilon_{k}\left(t\right)$. Such results are intuitively expected and are fully compatible with our classical notions of work and heat, in~agreement with the strategy adopted while proposing the gauge-invariant theory of quantum thermodynamics. This does not imply that the coherences play no role in the transfer of energy. The~generation of coherences are related to the rate at which energy is being transferred to the system in the form of work and/or heat, thus affecting all thermodynamic processes, including work extraction and the efficiency of heat engines. We will come back to this discussion while presenting some examples in the next~section.

%%%%%%%%%%%%%%%%%%%%%%%%%%%%%%%%%%%%%%%%%%%%%%%%%
%%%%%%%%%%%%%%%%%%%%%%%%%%%%%%%%%%%%%%%%%%%%%%%%%
%%%%%%%%%%%%%%%%%%%%%%%%%%%%%%%%%%%%%%%%%%%%%%%%%
\section{Applications} 
\label{sec:examples}

As explained in the last section, deviations between the two approaches appear together with quantum effects. This can be seen in the examples discussed in this section, where the role of coherences is being~exposed.

Let us first discuss a single qubit, described by the Hamiltonian $H_{t} = \alpha_{t}\sigma_{z}$, undergoing a non-unitary evolution described by the generalized amplitude damping or the phase damping channels. These processes can be described by Lindblad master equations of the general form 
\[
\dot \rho_t = -i\comm{H_{t}}{\rho_t} + \mathcal{D}[\rho_t],
\]
with $\mathcal{D}$ being the non-unitary part of the dynamics (dissipator).  

For the dephasing channel the dissipator is given by
\begin{equation}
    \mathcal{D}[\rho] = -\frac{\Gamma_{\mathrm{dec}}}{2}\comm{\sigma_z}{\comm{\sigma_{z}}{\rho}},
\end{equation} 
with $\sigma_s$ being the $s$-th Pauli matrix, while $\Gamma_{\mathrm{dec}}$ represents the decoherence rate. The~generalized amplitude damping is described by the dissipator
\begin{eqnarray}
  \mathcal{D}[\rho] &=& \Gamma_a(\bar{n}+1)\left[\sigma_{-}\rho \sigma_{+} - \frac{1}{2}\left\lbrace \sigma_{+}\sigma_{-},\rho\right\rbrace\right] \nonumber \\
  &+& \Gamma_a\bar{n}\left[\sigma_{+}\rho \sigma_{-} - \frac{1}{2}\left\lbrace \sigma_{-}\sigma_{+},\rho\right\rbrace\right],
\end{eqnarray}
where $\sigma_{+}$ and $\sigma_{-}$ are the usual spin ladder operators while $\bar{n} = \left(e^{\beta \omega} - 1\right)^{-1}$ stands for the mean excitation number of the bath mode with frequency $\omega$. $\beta$ stands for the inverse temperature while $\Gamma_a$ is the decoherence~rate. 

Since the Hamiltonian is already diagonal we take $u_{t} = \mathds{1}$, implying that $\dot{h}_{t} = \dot{\alpha}_t\sigma_z$. This leads to the following result for the invariant work
\begin{equation}
    W_{\mathrm{inv}} = \int_{0}^{\tau}\dd t\dot{\alpha}_t\expval{\frac{\partial H_t}{\partial \alpha_t}},
    \label{eq:work}
\end{equation}
for both channels. Of~course, $W_{\mathrm{inv}} = 0$ if $\alpha_t$ is time independent. Note that the above calculations do not assume any specific form of the driving. What matters is the power, i.e, the~rate of energy transfer which is, as~expected, influenced by the generation of~coherences. 

Let us now move to the invariant heat. Since $u_{t}$ is time independent, we have $Q_{\mathrm{inv}} = Q_{u}$, resulting in\vspace{6pt}
\begin{equation}
    Q_{\mathrm{inv}} = \int_{0}^{\tau}\dd t \tr{\mathcal{D}[\rho_t]H_{t}}.
    \label{eq:heat}
\end{equation}

For the case of the dephasing channel we obtain $Q_{\mathrm{inv}}^{\mathrm{deph}} = 0$. There is no heat in this case due to the fact that the coherences are destroyed by the decoherence process induced by the action of the bath. The~final state is diagonal in the energy eigenbasis and there is no energy flux from the system to the environment (the interaction Hamiltonian is diagonal). Therefore, all the energy transferred to the system by the driving must be in the form or~work. 

In the case of the generalized amplitude damping the invariant heat takes the form $Q_{\mathrm{inv}}^{\mathrm{gad}} = \int_{0}^{\tau}\dd t \Phi_{E}^{t}$, with~$\Phi_{E}^{t}$ being the instantaneous energy flux. In~this case, the~final state of the system is also diagonal. However, beyond~the decoherence process induced by the bath, there is also an energy exchange between the system and the environment. This energy is the~heat.

Since the dephasing channel only involves the exchange of phase information between the system and the environment, while the amplitude damping is characterized by energy exchange, the~above results agree with what is expected from classical thermodynamics. It is therefore important to stress that our gauge-invariant approach agrees with the standard one in cases which are consistently described from a point of view of classical thermodynamics, but~may differ when this is not the~case.

Let us now consider the unitary evolution of an externally driven single qubit whose Hamiltonian can be written as $H_{t} = \sigma_{z} + \gamma_{t}\sigma_{x}$, where $\gamma_t$ is a time-dependent driven frequency. In~this case we will have coherences being generated during the evolution. In~the appendix we show that the invariant heat is directly linked with the generations of coherences while the invariant work comes from the changes in the eigenenergies of the system. Figure~\ref{fig:heat_work_coh} shows   (see Appendix \ref{app:examples} for details) the invariant heat and work, along with the relative entropy of coherence. The~latter quantity is a well-behaving measure of quantum coherence defined as~\cite{Baumgratz2014}
\begin{equation}
    C = S(\rho_{\mathrm{diag}}) - S(\rho),
\end{equation}
where $\rho_\mathrm{diag}$ is the density operator with the off-diagonal (in the the energy eigenbasis) elements removed, while $S$ stands for the von Neumann~entropy.

Note that this example is in deep contrast with the usual understanding met in quantum thermodynamics, where it is assumed that a unitary evolution does not involve heat~\cite{Alicki1979,Kieu2004,Binder2018,Landi2020}. However, as~mentioned, this is in perfect agreement with the notion of heat introduced for closed systems and the definition of quantum adiabatic processes~\cite{Bunin2011,Polkovnikov2008}. Therefore, our results provide a firm physical ground---the gauge invariance principle---upon which we can justify why heat in closed quantum systems is linked with the delocalization of energy, explaining its oscillations for a closed evolution, where coherences change in~time. 

As our last example, let us consider a driven LMG model~\cite{Lipkin1965}
\begin{equation}
H = \frac{\lambda(t)}{j}J_{x}^{2} - J_{z},
\end{equation}
where $\lambda(t) = (1/4)\left(\tanh\left[\alpha(t-1)\right] + \tanh[\alpha]\right)$, $\alpha$ is a constant while $J_{i}$ denotes the $i$-th component of the angular momentum operator, whose value is $j$ [so the dimension of the associated Hilbert space is $(2j+1)$]. This choice has two advantages. First, the~intensity of the squeezing parameter $\lambda$ is in the interval $[0,0.5]$, which assures that the energy spectrum is not degenerated and we also avoid the critical points of the model. Secondly, we can consider situations from a very slow driving, when $\alpha \ll 1$, to~sudden quenches, when $\alpha\rightarrow\infty$. The~details of the numerical calculations are given in Appendix \ref{app:examples} and in Figure~\ref{fig:lmg_alpha} we present the invariant heat as well as the quantum coherence, considering distinct values of $\alpha$. 

%%%%%%%%%%%%%%%%%%%%%%%%%%%%%%%%%%%%%%%%%%%%%%%%%%%%%%%%%%%%%%%%%%%%%%%%
\begin{figure}[h]
    %\centering
    \includegraphics[width=0.5\textwidth]{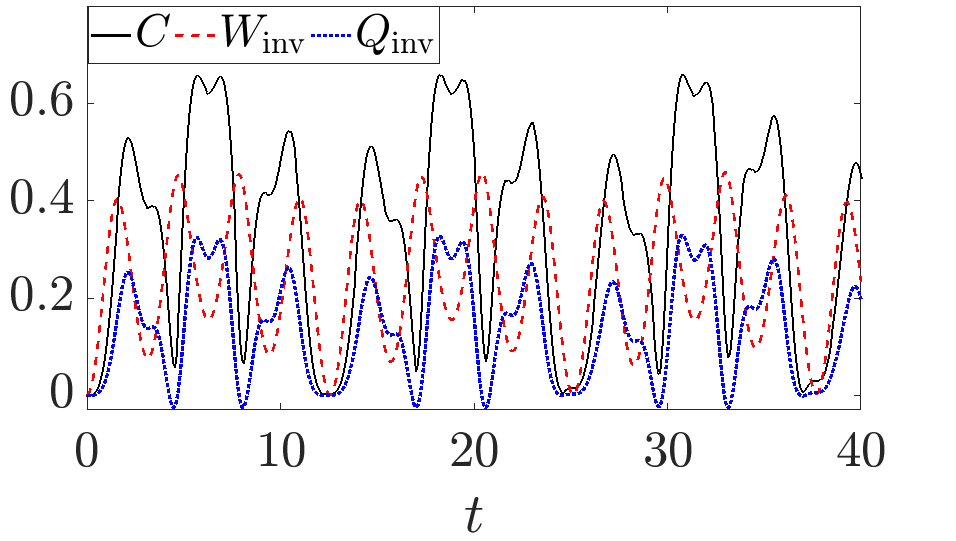}
    \caption{Energies and coherence. The~red dashed line represents the invariant work $W_{\mathrm{inv}}$ while the solid black line is the coherence and the dotted blue line is the invariant heat. We considered $\gamma_t = \cos(t)$ and we start from the ground state of the initial Hamiltonian. It is clear that the heat dynamics are perfectly correlated with the coherence, as~can be seen from the calculations shown in Appendix \ref{app:examples}.}
    \label{fig:heat_work_coh}
\end{figure}
%%%%%%%%%%%%%%%%%%%%%%%%%%%%%%%%%%%%%%%%%%%%%%%%%%%%%%%%%%%%%%%%%%%%%%%%

\vspace{-12pt}
%%%%%%%%%%%%%%%%%%%%%%%%%%%%%%%%%%%%%%%%%%%%%%%%%%%%%%%%%%%%%%%%%%%%%%%%
\begin{figure}[h]
  %  \centering
    \includegraphics[width=0.48\textwidth]{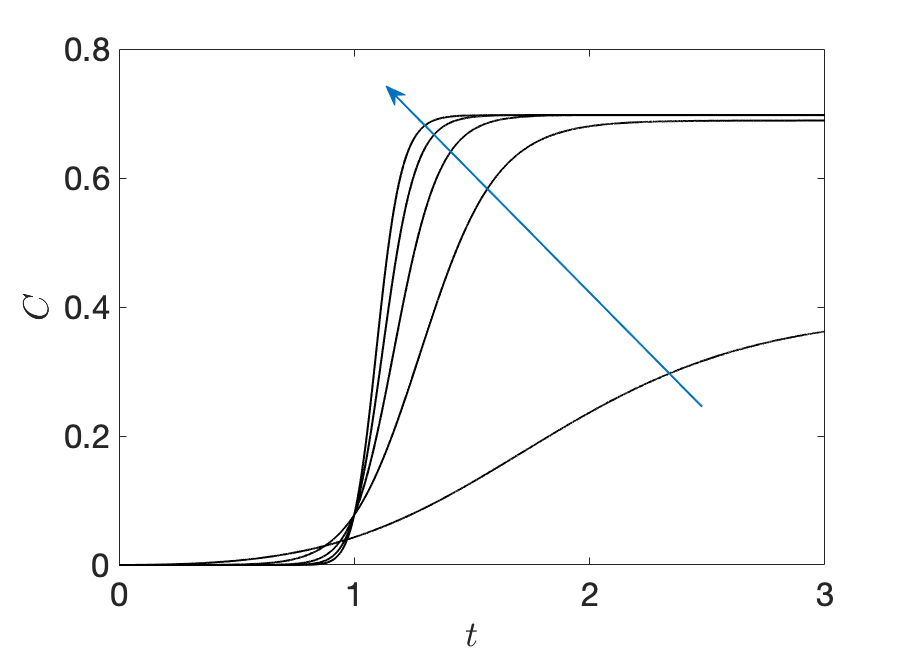} 
    \includegraphics[width=0.48\textwidth]{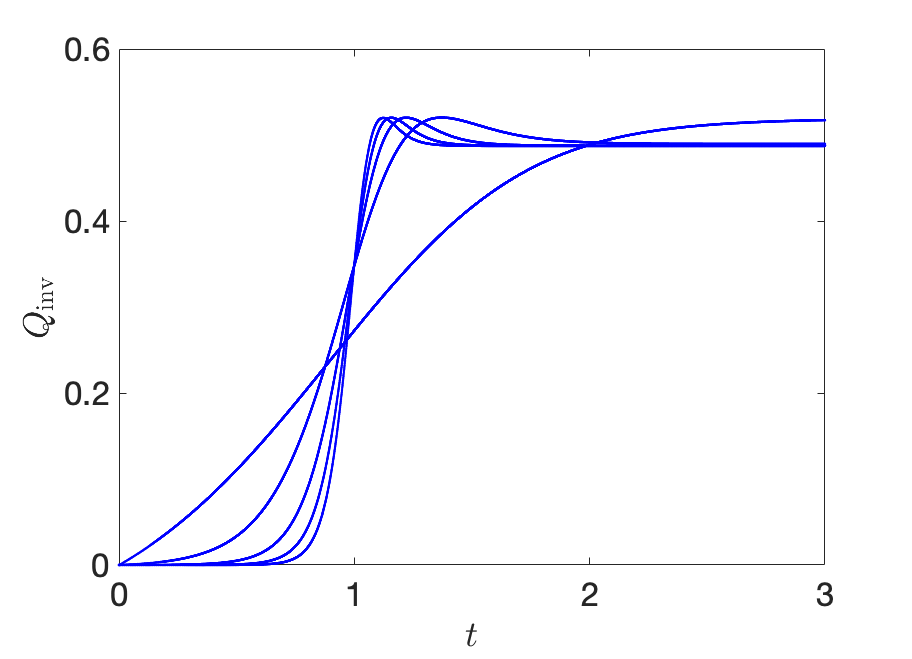}
    \caption{Heat and coherence. From~left to right we show the relative entropy of coherence $C$ and the invariant heat $Q_{\mathrm{inv}}$ as functions of time, for~distinct values of the speed of the protocol $\alpha$. The~blue arrow in the top panel shows the direction in which $\alpha$ is increasing, from~$\alpha=1$ to $\alpha=10$. We choose $j=50$ for all curves and the starting point was the energy ground state for $t=0$.}
    \label{fig:lmg_alpha}
\end{figure}
%%%%%%%%%%%%%%%%%%%%%%%%%%%%%%%%%%%%%%%%%%%%%%%%%%%%%%%%%%%%%%%%%%%%%%%%

This example shows how the heat scales with the generation of coherences. As~we can see, as~time passes, we transition from an adiabatic transformation of the Hamiltonian (too slow to create coherences) towards a regime where coherences are indeed generated and we also observe the increase in the generation of heat. In~this regime, transitions between energy eigenstates are inevitably induced by the fast changes in the Hamiltonian and such transitions are associated with~heat. 

Beyond the fundamental relevance, the~above results are important for the thermodynamic analysis of heat engines. For~instance, let us consider an Otto cycle consisting of four stages, two non-unitaries (equilibration with thermal baths) and two controlled unitary transformations. In~general, heat is associated with the non-unitary dynamics while all the energy transferred to/from the system during the unitary transformations is regarded as work. Since the efficiency is defined as the ratio between the extracted work and the heat absorbed from the hot bath, under~the present approach, where part of the energy during the unitary stages is heat, the~efficiency of the engine will decrease as we increase the speed at which these transformations are performed (see Figure~\ref{fig:lmg_alpha}). This is in complete agreement with the expected behavior of a heat engine, since fast processes take us far from equilibrium and, thus, more entropy must be produced and the engine efficiency must decrease, in~accord with the second law of thermodynamics. Our theory naturally encompasses such~behavior. 

%%%%%%%%%%%%%%%%%%%%%%%%%%%%%%%%%%%%%%%%%%%%%%%%%
%%%%%%%%%%%%%%%%%%%%%%%%%%%%%%%%%%%%%%%%%%%%%%%%%
%%%%%%%%%%%%%%%%%%%%%%%%%%%%%%%%%%%%%%%%%%%%%%%%%
\section{Discussion} 
\label{sec:discussions}

Concerning the first law of thermodynamics, the~usual notions of heat and work are based on the Clausius theorem. Heat is associated with the change in the thermodynamic entropy while work is linked with the changes in some externally controlled parameter. Such notions cannot be simply extended to the quantum world, thus leading to several definitions of work and heat that are, in~general, nonequivalent, but~operationally well-defined~\cite{Talkner2016}. The~theory proposed here leads to gauge-invariant definitions of thermodynamic quantities, in~particular, heat and work that are not solely related with the change in the information entropy. Work is associated with the eigenvalues of the Hamiltonian while heat is connected with the change in the eigenbasis of the Hamiltonian, thus being deeply linked with energy delocalization. This implies that we can have heat even in a closed quantum system, which is in sharp contrast with the usual notions of quantum heat~\cite{Alicki1979,Kieu2004,Binder2018,Landi2020}. However, we must remember that such a notion is well-understood in classical physics, where heat is associated with energy transferred to the degrees of freedom that are averaged out due to the coarse-grained nature of macroscopic measurements~\cite{Callen1985}. In quantum physics, things are trickier and, when considering closed systems, heat is usually associated with transitions (generation of coherences) in the energy eigenbasis~\cite{Bunin2011,Polkovnikov2008}. In the adopted perspective, coherences in the energy eigenbasis indicate that the energy variance does not vanish and this uncertainty is here interpreted as heat. Therefore, our work shows that this notion naturally emerges from the framework of gauge invariance, a~physical principle lying in the basis of all modern~physics. 

Regarding the notion of statistical work (and heat), the~most considered definition is based on the two-point measurement~\cite{Talkner2007,Campisi2011}. This protocol has been considered both theoretically~\cite{Esposito2009,Campisi2011,Landi2020} and experimentally~\cite{Batalhao2014,Zhang2015,Cerisola2017,Gomez2020,Zanin2019,Ribeiro2020,Peterson2016} in distinct contexts within quantum thermodynamics. Although~widely accepted, such scheme does not take into account the coherences, which are destroyed by the measurements. Therefore, statistical work is gauge-invariant by construction. Fully quantum fluctuation theorems were developed in order to take into account this contribution~\cite{Jevtic2015,Alhambra2016,Aberg2018,Santos2019,Francica2019,Kwon2019}. A~naturally arising question is to scrutinize or generalize/modify the fluctuation theorems from the perspective of the invariant work and heat presented in this work. The~implications of such definitions for the irreversible entropy production will certainly deepen our understanding of quantum coherence in non-equilibrium~thermodynamics.

Another question that can be addressed under this theory is the local thermalization of closed quantum many-body systems and how this is linked with the eigenstate thermalization hypothesis~\cite{Deutsch2018} and irreversible entropy production~\cite{Landi2020}.

Perhaps the most important and broad message of the current work is relevant for all aspects of quantum technologies associated with operational principles behind potential quantum engines, batteries or refrigerators. Our approach might in the future allow for the delineation of the boundaries between controlled quantum systems and their thermodynamic behavior, forming a bridge between classical and quantum thermodynamics, which starts in the latter paradigm and expands towards the former~one.

%%%%%%%%%%%%%%%%%%%%%%%%%%%%%%%%%%%%%%%%
%%%%%%%%%%%%%%%%%%%%%%%%%%%%%%%%%%%%%%%%
%%%%%%%%%%%%%%%%%%%%%%%%%%%%%%%%%%%%%%%%
\appendix

\section{Gauge Invariance of~Work}
\label{app:invariance}

We first present a formal proof that $W_{\mathrm{inv}}$ defined in Equation~\eqref{eq:inv_work} in the main text is invariant with respect to the emergent gauge transformations. To~this end we study
\begin{equation}
W_{\mathrm{inv}}[V_t\rho_tV_t^\dagger]=\int_{0}^{\tau}dt\mathrm{Tr}\left[V_t\rho_{t}V_t^\dagger u_{t}\dot{h}_{t}u_{t}^{\dagger}\right].
\end{equation}
Substituting the explicit form of $V_t$ and $h_t$ we find
\begin{equation}
\mathrm{Tr}\left[V_t\rho_{t}V_t^\dagger u_{t}\dot{h}_{t}u_{t}^{\dagger}\right]=\mathrm{Tr}\left[\rho_{t} u_{t}\left(\bigoplus_{k=1}^{p}v_{k}^\dagger \dot{\varepsilon}_{k}(t)v_{k}\right)u_{t}^{\dagger}\right].
\end{equation}
Since $v_{k}^\dagger \dot{\varepsilon}_{k}(t)v_{k}=\dot{\varepsilon}_{k}(t) \mathds{1}_{n_{k}}$, we obtain the desired~result.

In the next step we pass to the second assertion, namely, we study the invariant work from the perspective of group averaging. To~this end, we insert Equation~\eqref{Work not GI} in the Definition~\ref{def2}, obtaining
\begin{equation}
\int d\mathcal{G}W_u\left[V_{t}\rho_{t}V_{t}^{\dagger}\right]=\int_{0}^{\tau}dt\textrm{Tr}\left[\rho_{t}u_{t}f_{t}u_{t}^{\dagger}\right],
\end{equation}
where
\begin{equation}
f_{t}=\int d\mathcal{G}\left(\bigoplus_{k=1}^{p}v_{k}^{\dagger}\right)u_{t}^{\dagger}\dot{H}_{t}u_{t}\left(\bigoplus_{k=1}^{p}v_{k}\right).
\end{equation}
Further we expand
\begin{eqnarray*}
u_{t}^{\dagger}\dot{H}_{t}u_{t} & = & \dot{h}_{t}+h_{t}\dot{u}_{t}^{\dagger}u_{t}+u_{t}^{\dagger}\dot{u}_{t}h_{t}\\
 & = & \dot{h}_{t}+\left[h_{t},\dot{u}_{t}^{\dagger}u_{t}\right],
\end{eqnarray*}
where the last line follows from unitarity of $u_{t}$, which implies
$u_{t}^{\dagger}\dot{u}_{t}=-\dot{u}_{t}^{\dagger}u_{t}$. Therefore,
using the arguments spelled out previously, we find
\begin{equation}
f_{t}=\dot{h}_{t}+\left[h_{t},b_{t}\right],
\end{equation}
where
\begin{equation}
b_{t}=\int d\mathcal{G}\left(\bigoplus_{k=1}^{p}v_{k}^{\dagger}\right)\dot{u}_{t}^{\dagger}u_{t}\left(\bigoplus_{k=1}^{p}v_{k}\right).
\end{equation}
Due to the fundamental property of unitary group averaging we know
that $b_{t}=\bigoplus_{k=1}^{p}b_{k}\left(t\right)\mathds{1}_{n_{k}}$, for~some functions $b_k(t)$. Therefore, we find that $[h_t,b_t]=0$, so the group averaged work is the same as the invariant work proposed in Theorem~\ref{th1}.

Finally, let us consider the heat and its connections to quantum coherence. In the energy eigenbasis the coherent heat Equation~\eqref{CH} reads
\begin{align}
Q_{c}=\int_{0}^{\tau}dt\sum_{j l}c_{jl}(t)\left[
\varepsilon_{j}\left(t\right) \left\langle a_{l}\left(t\right)\left|\dot{a}_{j}\left(t\right)\right\rangle\right.\right.\\ \left.+\varepsilon_{l}\left(t\right) \left\langle \dot{a}_{l}\left(t\right)\left|a_{j}\left(t\right)\right\rangle\right.\right].
\end{align}
Since for all times we have
\begin{equation}
\left\langle a_{l}\left(t\right)\left|a_{j}\left(t\right)\right\rangle \right.=\delta_{lj},
\end{equation}
we also consequently obtain
\begin{equation}
\left\langle a_{l}\left(t\right)\left|\dot{a}_{j}\left(t\right)\right\rangle \right.+\left\langle \dot{a}_{l}\left(t\right)\left|a_{j}\left(t\right)\right\rangle \right.=0.
\end{equation}
Therefore, we obtain the final result
\begin{equation}
Q_{c}=\int_{0}^{\tau}dt\sum_{j l}c_{jl}\left(t\right)\left\langle \dot{a}_{l}\left(t\right)\left|a_{j}\left(t\right)\right\rangle \right.\left(\varepsilon_{l}\left(t\right)-\varepsilon_{j}\left(t\right)\right).
\end{equation}
Clearly, only off-diagonal  terms of the density matrix (coherences) contribute to this part of~heat. 

%%%%%%%%%%%%%%%%%%%%%%%%%%%%%%%%%%%%%%%%%%%%%%%%%%%%%%%%%%%%%%%%%%%%%%%%%%%%%
%%%%%%%%%%%%%%%%%%%%%%%%%%%%%%%%%%%%%%%%%%%%%%%%%%%%%%%%%%%%%%%%%%%%%%%%%%%%%
%%%%%%%%%%%%%%%%%%%%%%%%%%%%%%%%%%%%%%%%%%%%%%%%%%%%%%%%%%%%%%%%%%%%%%%%%%%%%
\section{Applications}
\label{app:examples}

Let us consider the unitary dynamics of a single qubit governed by the Hamiltonian
\begin{equation}
    H_{t} = \sigma_{z} + \gamma_{t}\sigma_{x}.
\end{equation}
This Hamiltonian can be put into a diagonal form $h_{t} = u_{t}^{\dagger}H_{t}u_{t} = \mbox{diag}(-\lambda_t,\lambda_t)$, with~$\lambda_{t}^{2} = 1 + \gamma_{t}^{2}$. Here, $u_t$ is the matrix whose columns are the normalized eigenvectors of $H_{t}$
\begin{equation}
u_t = 
\begin{bmatrix}
\frac{1 - \lambda_{t}}{n_{-}} & \frac{1 + \lambda_{t}}{n_{+}}\\
\frac{\gamma_{t}}{n_{-}} & \frac{\gamma_{t}}{n_{+}}
\end{bmatrix},
\end{equation}
with $n_{\pm} = \sqrt{2\lambda_{t}(\lambda_{t} \pm 1)}$. Denoting $\left\lbrace \ket{a_{1}},\ket{a_{2}}\right\rbrace$ the eigenstates of the Hamiltonian, one can show that
\begin{equation}
    \braket{\dot{a}_{2}}{a_{1}} \equiv - \braket{\dot{a}_{1}}{a_{2}} = \frac{\dot{\gamma}_{t}}{2\lambda_{t}^{2}} \hspace{0.2cm} \mbox{and} \hspace{0.2cm} \braket{\dot{a}_{i}}{a_{i}} = 0.
\end{equation}

The invariant heat can be written as
\begin{equation}
Q_{\mathrm{inv}} = 2\int_{0}^{\tau}\dd t \mbox{Re}[c_{12}(t)]\frac{\dot{\gamma}_{t}}{\lambda_{t}}.
\end{equation}
Note that the only contribution that we have here comes from the coherences of the density operator in the energy eigenbasis. However, it is important to stress that, if~the eigenbasis is constant, the~coherences does not matter and the heat is zero. The~important thing here is the change in the eigenbasis of the~Hamiltonian.

The invariant work takes the form
\begin{equation}
    W_{\mathrm{inv}} = \int_{0}^{\tau}\dd t [ c_{22}(t) - c_{11}(t)]\dot{\lambda}_{t}.
\end{equation}

We shall numerically solve the evolution equation for the density operator (for notation brevity we omit time dependence of $c_{ij}$ coefficients)
\begin{equation}
    \dv{\rho_t}{t} = -i\comm{H_{t}}{\rho_t} = -i\sum_{i,j}\comm{H_{t}}{c_{ij}\dyad{a_{i}}{a_{j}}}.
\end{equation}
Since
\begin{equation}
    \dv{\rho_t}{t} =  \sum_{i,j}\left[\dot{c}_{ij}\dyad{a_{i}}{a_{j}} + c_{ij}\dyad{\dot{a}_{i}}{a_{j}} + c_{ij}\dyad{a_{i}}{\dot{a}_{j}}\right],
\end{equation}
one can in a straightforward way show that the coefficients $c_{ij}$ fulfill the following set of differential equations
\begin{eqnarray}
\dot{c}_{11} &=& - \frac{\dot{\gamma}_{t}}{2\lambda_{t}^{2}}\left(c_{12} + c_{21}\right), \nonumber \\
\dot{c}_{12} &=& 2i\lambda_{t}c_{12} + \frac{\dot{\gamma}_{t}}{2\lambda_{t}^{2}}\left(c_{11} - c_{22}\right), \nonumber \\
\dot{c}_{21} &=& - 2i\lambda_{t}c_{21} - \frac{\dot{\gamma}_{t}}{2\lambda_{t}^{2}}\left(c_{22} - c_{11}\right), \nonumber \\
\dot{c}_{22} &=& \frac{\dot{\gamma}_{t}}{2\lambda_{t}^{2}}\left(c_{12} + c_{21}\right).
\end{eqnarray}

Let us move now to the LMG model. The~Schr\"{o}dinger equation reads $i\dot{\ket{\psi}} = H\ket{\psi}$. We can expand the state $\ket{\psi}$ at any time in the basis defined by the relations $J^{2}\ket{j,m} = j(j+1)\ket{j,m}$ and $J_{z}\ket{j,m} = m\ket{j,m}$ in the form
\[
\ket{\psi} = \sum_{m=-j}^{j}c_{m}(t)\ket{j,m}.
\]
By inserting this expansion in the Schr\"{o}dinger equation and projecting the result into the bra $\bra{n}$ we obtain the following set of differential equations for the expansion coefficients
\begin{widetext}
\begin{eqnarray}
i\dv{c_{n}(t)}{t} &=& \left[\frac{j(j+1)-n^{2}}{2}-n\right]c_{n}(t) \nonumber\\
&+& \frac{\lambda(t)}{4}\sqrt{(j+n-1)(j+n)(j-n+1)(j-n+2)}c_{n-2}(t) \nonumber\\
&+& \frac{\lambda(t)}{4}\sqrt{(j+n+1)(j-n)(j-n-1)(j+n+2)}c_{n+2}(t),
\end{eqnarray} 
\end{widetext}
which can be numerically solved. From~this we can reconstruct the density operator of the system at any instant of~time. 

The next thing we calculate is the time derivative of the diagonal form of the Hamiltonian. This is easily performed by numerically diagonalizing the full Hamiltonian at each instant of time and computing the time derivative of the eigenvalues. We do not actually need to compute the time derivative of the unitary operator $u_{t}$ (that diagonalizes the Hamiltonian), since we can write
\begin{eqnarray}
\dv{U}{t} &=& \dv{\lambda(t)}{t}\mbox{Tr}\left[\pdv{H}{\lambda}\rho_{t}\right] \nonumber\\ 
&=&\frac{\alpha}{j}\sech^{2}\left[\alpha(t-1)\right]\mbox{Tr}\left[J_{x}^{2}\rho_{t}\right].
\end{eqnarray}
Now, since we have
\begin{equation}
\dv{U}{t} = \dv{W_{\mathrm{inv}}}{t} + \dv{Q_{\mathrm{inv}}}{t},
\end{equation}
which is the first law, and~\begin{equation}
\dv{W_{\mathrm{inv}}}{t} = \mbox{Tr}\left[u_{t}\dv{h}{t}u_{t}^{\dagger}\rho_{t}\right],
\end{equation}
we can obtain the heat by integrating the last two equations. This will result in the coherent heat since the system is closed and $Q_{\mathrm{inv}} = Q_{\mathrm{c}}$.

%----------------------------------
\begin{acknowledgments}
The authors are in debit to Gabriel Landi for his critical reading of the~manuscript. \L R and LCC acknowledge support by the Foundation for Polish Science (IRAP project, ICTQT, contract no. 2018/MAB/5, co-financed by EU within Smart 447 Growth Operational Programme). LCC acknowledges support by the National Institute for the Science and Technology of Quantum Information (INCT-IQ), Grant No. 465469/2014-0, by the National Council for Scientific and Technological Development (CNPq), Grant No 308065/2022-0, and by Coordination of Superior Level Staff Improvement (CAPES).
\end{acknowledgments}

%%%%%%%%%%%%%%%%%%%%%%%%%
%%%%%%%%%%%%%%%%%%%%%%%%%
%%%%%%%%%%%%%%%%%%%%%%%%%

\bibliographystyle{apsrev4-1}

\end{document}